\documentclass[12pt]{article}
\usepackage{graphicx}
\usepackage{graphics}
\usepackage{amsthm}
\usepackage{amssymb, amsmath}
\title{Mellin transforms with only critical zeros:  Chebyshev and Gegenbauer functions}
\author{Mark W. Coffey\\
Department of Physics\\
Colorado School of Mines\\
Golden, CO  80401\\
USA\\
mcoffey@mines.edu\\
Matthew C. Lettington\\
School of Mathematics\\
Cardiff University\\
P. O. Box 926\\
Cardiff CP24 4AG\\
UK\\
LettingtonMC@cf.ac.uk}
\date{August 27, 2013}
\begin{document}
\maketitle
\baselineskip=25 pt
\begin{abstract}

We consider the (generalized) Mellin transforms of certain Chebyshev functions based upon the Chebyshev polynomials.  We show that the transforms have polynomial factors whose zeros lie all on the critical line or on the real line.  The polynomials with zeros only on the critical
line are identified in terms of certain $_3F_2(1)$ hypergeometric functions.  
Furthermore, we extend this result to a $1$-parameter family of polynomials with zeros
only on the critical line.  These polynomials possess the functional equation $p_n(s;\beta)=(-1)^{\lfloor n/2 \rfloor} p_n(1-s;\beta)$.  We then present the generalization
to the Mellin transform of certain Gegenbauer functions.
The results should be of interest to special function theory, combinatorics, and analytic number theory. 

\end{abstract}
 
\vspace{.25cm}
\baselineskip=15pt
\centerline{\bf Key words and phrases}
\medskip 
Mellin transformation, Chebyshev polynomials, Gegenbauer polynomial, hypergeometric function, critical line, zeros, functional equation

\bigskip
\noindent
{\bf 2010 MSC numbers}
\newline{33C20, 33C45, 42C05, 44A20, 30D05}  

\baselineskip=25pt

\pagebreak
\centerline{\bf Introduction}

\medskip

In a series of papers, we are considering certain Mellin transforms comprised of classical orthogonal polynomials that yield polynomial factors with zeros only on the critical line Re $s=1/2$ or else only on the real axis.  Such polynomials have many important applications to
analytic number theory, in a sense extending the Riemann hypothesis.  For example,
using the Mellin transforms of Hermite functions, Hermite polynomials multiplied by a
Gaussian factor, Bump and Ng \cite{bumpng} were able to generalize
Riemann's second proof of the functional equation of the zeta function $\zeta(s)$, and to
obtain a new representation for it.  The polynomial factors turn out to be certain 
$_2F_1(2)$ Gauss hypergeometric functions \cite{coffeymellin}.  

In a different setting, the polynomials $p_n(x)= ~_2F_1(-n,-x;1;2)=(-1)^n ~_2F_1(-n,x+1;1;2)$ and $q_n(x)=i^n n! p_n(-1/2-ix/2)$ were studied \cite{kirsch}, and they directly correspond to the Bump and Ng polynomials with $s=-x$.  Kirschenhofer, Peth\"{o}, and Tichy considered combinatorial properties of $p_n$, and developed Diophantine properties of them.  Their analytic results for $p_n$ include univariate and bivariate generating functions, and that its
zeros are simple, lie on the line $x=-1/2+it$, $t \in \mathbb{R}$, and that its zeros
interlace with those of $p_{n+1}$ on this line.  We may observe that these polynomials
may as well be written as $p_n(x)={{n+x} \choose n} ~_2F_1(-n,-x;-n-x;-1)$, or
$$p_n(x)={{(-1)^n 2^n \Gamma(n-x)} \over {n!\Gamma(-x)}} ~_2F_1\left(-n,-n;x+1-n;{1 \over 2}
\right),$$
where $\Gamma$ is the Gamma function.

The Hermite polynomials being certain cases proportional to Laguerre polynomials $x^\delta L_n^{\pm 1/2}(x^2)$, $\delta=0$ or $1$, the generalization to Mellin transforms of Laguerre functions has been made \cite{bumpchoi,coffeymellin} and now the polynomial factors are a family of other $_2F_1(2)$ functions.  The Laguerre functions are ${\cal L}_n^\alpha(x) =x^{\alpha/2}e^{-x/2}L_n^\alpha(x)$, for $\alpha>-1$, and their Mellin transforms are of the
form $M_n^\alpha(s)=2^{s+\alpha/2}\Gamma(s+\alpha/2)P_n^\alpha(s)$.  Mixed recursion
relations are known for the polynomials $P_n^\alpha$, as well as a generating function, and
they satisfy the functional equation $P_n^\alpha(s)=(-1)^n P_n^\alpha(1-s)$.  Separately one
of us has addressed the Mellin transform of certain generalized Hermite functions 
\cite{coffeygen}.  The polynomial factors have zeros only on the critical line and possess
a reciprocity relation.

In this article, we address the Mellin transforms of certain Chebyshev and Gegenbauer functions, and are able to identify the resulting polynomial factors in terms of certain generalized hypergeometric functions $_3F_2(1)$.  The key result is that these polynomials possess zeros only 
on the critical line, or on the real line. 

We use standard notation.
Let $_pF_q$ be the generalized hypergeometric function, $(a)_n=\Gamma(a+n)/\Gamma(a)=(-1)^n {{\Gamma(1-a)} \over {\Gamma(1-a-n)}}$ the Pochhammer symbol, 
and $B(x,y)=\Gamma(x)\Gamma(y)/\Gamma(x+y)$ the Beta function.  

Further, let $T_n$ and $U_n$ be the Chebyshev polynomials of the first and
second kinds, respectively \cite{mason,rivlin}.  
The Chebyshev polynomials figure prominently in several areas such as approximation theory
and other parts of numerical analysis.  An abundance of analytic properties of these 
polynomials is known.  In addition, some of their divisibility properties have been studied.
While Chebyshev polynomials $V_n$ and $W_n$ of the third and fourth kind may be introduced,
they will not be examined in this paper.

Treating the Mellin transforms of Chebyshev functions suffices for otherwise also treating
analogous functions based upon the Pell or Morgan-Voyce polynomials that are scaled or 
shifted forms of the Chebyshev functions.  The Pell polynomials $\tilde{p}_n$ are variously defined.
If we take them as given by the generating function
$${{xt} \over {1-2xt-t^2}}=\sum_{m=1} \tilde{p}_m(x)t^m,$$
then they correspond to $\tilde{p}_n(x)=x(-i)^{n-1}U_{n-1}(ix)$ for $n \geq 1$.  For the
Morgan-Voyce polynomials \cite{swamy} we have $b_n(x)=U_n(x/2+1)-U_{n-1}(x)$ and
$B_n(x)=U_n(x/2+1)$ for $n \geq 1$ and $b_0=B_0=1$.

A by-product of \cite{bumpchoi} is a reciprocity law for the polynomial factors, relating 
values at negative integers.  In particular, the $\alpha=0$ case reads $P_n^{(0)}(-m)=
P_m^{(0)}(-n)$.  Interestingly, $P_n^{(0)}(-m)$ has a combinatorial interpretation as
the number of lattice points $(x_1,\ldots,x_n) \in \mathbb{Z}^n$ such that $\sum_i |x_i|\leq m$.  This points to other possible connections with combinatorics.

In fact, polynomials with only real zeros commonly arise in combinatorics and elsewhere.
Two examples are Bell polynomials \cite{harper} and Eulerian polynomials \cite{comtet} (p. 292).  This suggests that there may be other relations of our results to discrete
mathematics and other areas.  In this regard, we mention matching polynomials with only
real zeros.  The matching polynomial $M(G,x)=\sum_k (-1)^k p(G,k)x^k$ counts the matchings
in a graph $G$.  Here, $p(G, k)$ is the number of matchings of size $k$, i.e., the
number of sets of $k$ edges of $G$, no two edges having a common vertex.
$M(G,x)$ satisfies recurrence relations and has only real zeros and $M(G - {v}, x)$
interlaces $M(G,x)$ for any $v$ in the vertex set of $G$ (e.g., \cite{godsil}).  As regards
\cite{bumpchoi,coffeymellin}, we note that 
the classical orthogonal polynomials are closely related to the matching polynomials. For
example, the Chebyshev polynomials of the first two kinds are the matching polynomials of paths and cycles respectively, and the Hermite polynomials and the Laguerre polynomials are 
the matching polynomials of complete graphs and complete bipartite graphs, respectively.


In \cite{bumpng,coffeymellin,coffeyxi}, Mellin transforms were used on $[0,\infty)$.
Here we consider Mellin transformations for functions supported on $[0,1]$,
$$({\cal M}_0f)(s)=\int_0^1 f(x)x^s {{dx} \over x}.$$
For properties of the Mellin transform, we mention \cite{butzer}.

We put, for Re $s>0$,
$$M_n(s) \equiv \int_0^1 x^{s-1} U_n(x) {{dx} \over {(1-x^2)^{1/4}}}.  \eqno(1.1)$$
For $n$ odd, these transforms additionally hold for Re $s>-1$.
Let $p_n(s)$ denote the polynomial factor of $M_n(s)$.

In fact, we could equally well employ
$$({\cal M}_1f)(s)=\int_1^\infty f(x)x^s {{dx} \over x},$$
and $({\cal M}f)(s)=({\cal M}_0f)(s)+({\cal M}_1f)(s)$, where the latter representation
is to be taken as the analytic continuation of each term.  For what we present, it is
specifically the analytic continuation of the Gamma function to the whole complex plane
that permits $({\cal M}f)(s)$ to exist also through out $\mathbb{C}$.  Indeed, here the
contributions $({\cal M}_0f)(s)$ and $({\cal M}_1f)(s)$ are `companions'--the analytic
continuations of one another.  At the end of the final Discussion section, we further
describe the notion of generalized Mellin transform that is present. 


The Chebyshev polynomials $U_n(x)$ are the $\lambda=1$ case of the Gegenbauer polynomials
$C_n^\lambda(x)$ with $\lambda>-1/2$ (e.g., \cite{andrews}). The latter polynomials (e.g.,
\cite{rainville} Ch. 17) are orthogonal on $[-1,1]$ with weight function $(1-x^2)^{\lambda-1/2}$.  
The Gegenbauer polynomials, and thus the Chebyshev polynomials, satisfy ordinary differential
equations.  A series representation of the Gegenbauer polynomials is given by
$$C_n^\lambda(x)={{(2\lambda)_n} \over {n!}} ~_2F_1\left(2\lambda+n,-n;\lambda+{1 \over 2};
{{1-x} \over 2}\right),$$
immediately showing that $C_n^\lambda(1)=(2\lambda)_n/n!$.
In Proposition 5 we provide the appropriate generalization of (1.1) so that the resulting polynomial factors again have zeros only on the critical line.

{\bf Proposition 1}.
The polynomials $p_n$ satisfy the following recursion relation, with 
$p_0=\Gamma(3/4)/2$ and $p_1=\Gamma(3/4)$.
{\newline For $n$ even,}
$$p_n(s)=sp_{n-1}(s+1)-{1 \over 2}\left(s+n-{1 \over 2}\right)p_{n-2}(s), \eqno(1.2a)$$
and for $n$ odd,
$$p_n(s)=2p_{n-1}(s+1)-{1 \over 2}\left(s+n-{1 \over 2}\right)p_{n-2}(s). \eqno(1.2b)$$

{\bf Proposition 2}.
The polynomials $p_n(s)$, of degree $\lfloor n/2 \rfloor$, satisfy the functional equation
$p_n(s)=(-1)^{\lfloor n/2 \rfloor} p_n(1-s)$.  These polynomials have zeros only on the
critical line.  Further, all zeros $\neq 1/2$ occur in complex conjugate pairs.

{\bf Proposition 3}.  The polynomials 
$$p_n(s;\beta)={{\Gamma\left({{n+s} \over 2}\right)} \over {\Gamma\left({{s+\varepsilon} \over 2}\right)}}
~_3F_2\left(1-\beta,{{1-n} \over 2},-{n \over 2};2(1-\beta),1-{{(n+s)} \over 2};1\right),$$
wherein $\varepsilon=0$ for $n$ even and $=1$ for $n$ odd, $\beta<1$, 
of degree $\lfloor n/2 \rfloor$, satisfy the functional equation
$p_n(s;\beta)=(-1)^{\lfloor n/2 \rfloor} p_n(1-s;\beta)$.  These polynomials have zeros only 
on the critical line, and all zeros $\neq 1/2$ occur in complex conjugate pairs.

{\bf Corollary 1}.  (a) The properties of Proposition 3 are satisfied by the polynomials
$$p_n(s;0)={{2(n+s)} \over {(n+1)(n+2)}}{{\Gamma\left({{n+s} \over 2}\right)} \over {\Gamma\left({{s+\varepsilon} \over 2}\right)}} \left[1- ~_2F_1\left(-{{(n+1)} \over 2},-{n \over
2}-1;-{{(n+s)} \over 2};1\right)\right]$$
$$={{2(n+s)} \over {(n+1)(n+2)}}{{\Gamma\left({{n+s} \over 2}\right)} \over {\Gamma\left({{s+\varepsilon} \over 2}\right)}} \left[1- {{\Gamma\left(-{{(n+s)} \over 2}\right)} \over {\Gamma\left({{1-s} \over 2}\right)}} {{\Gamma\left({{n+3-s} \over 2}\right)} \over {\Gamma\left(1-{s \over 2}\right)}}\right].$$
(b) More generally, for $\beta$ a negative integer, the properties of Proposition 3 are satisfied 
by the polynomials $p_n(s;-m)$, and these polynomials may be written in terms of elementary
factors and the Gamma function.

There are many known transformations of $_3F_2$ functions, and, in particular, of terminating
$_3F_2(1)$ functions, as illustrated in the Appendix.  We present another hypergeometric
form of the Mellin transforms $M_n(s)$, and accordingly, of their polynomial factors.
\newline{\bf Proposition 4}.
$$M_{2n}(s)=M_0(s)(-1)^n ~_3F_2\left(-n,n+1,{s \over 2};{{2s+3} \over 4},{1 \over 2};1\right),$$
$$M_{2n+1}(s)=2M_0(s+1)(-1)^n (n+1)~_3F_2\left(-n,n+2,{{s+1} \over 2};{{2s+5} \over 4},{3 
\over 2};1\right),$$
wherein $M_0(s)=\Gamma(3/4)\Gamma(s/2)/[2\Gamma(s/2+3/4)]$.

{\bf Proposition 5}.  For $\lambda>-1/2$, let
$$M_n^\lambda(s)=\int_0^1 {{C_n^\lambda(x)x^{s-1}} \over {(1-x^2)^{3/4-\lambda/2}}}dx
=\int_0^{\pi/2} \cos^{s-1} \theta ~C_n^\lambda(\cos \theta) \sin^{\lambda-1/2} \theta ~d\theta, \eqno(1.3)$$
wherein we assume Re $s>0$ for $n$ even and Re $s>-1$ for $n$ odd.  Then  
(a) (recurrence)
$$nM_n^\lambda(s)=2(\lambda+n-1)M_{n-1}^\lambda(s+1)-(2\lambda+n-2)M_{n-2}^\lambda(s)$$
with
$$M_0^\lambda(s)={{\Gamma\left({\lambda \over 2}+{1 \over 4}\right)\Gamma\left({s \over 2}\right)}
\over {2\Gamma \left({{s+\lambda} \over 2}+{1 \over 4}\right)}},$$
and $M_1^\lambda(s)=2\lambda M_0^\lambda(s+1)$.
(b) (generating function)
$$G^\lambda(s,t) \equiv \sum_{k=0}^\infty M_k^\lambda(s)t^k =\int_0^1 {1 \over {(1-x^2)^{3/4-\lambda/2}}}{x^{s-1} \over {(1-2tx+t^2)^\lambda}}dx$$
$$={1 \over {(1+t^2)^\lambda}}{{\Gamma\left({1 \over 4}+{\lambda \over 2}\right)} \over 2}\left[{{\Gamma(\lambda)\Gamma\left({s \over 2}\right)} \over {\Gamma\left({{s+\lambda} \over 2}+{1 \over 4}\right)}} ~_3F_2\left({{\lambda+1} \over 2},{\lambda \over 2},{s \over 2};{1 \over 2},{{s+\lambda} \over 2}+{1 \over 4};{{4t^2} \over {(1+t^2)^2}}\right) \right.$$
$$\left. + {{2t\Gamma(\lambda+1)} \over {(1+t^2)}}{{\Gamma\left({{s+1} \over 2}\right)} \over {\Gamma\left({{s+\lambda} \over 2}+{3 \over 4}\right)}} ~_3F_2\left({{\lambda+1} \over 2},
1+{\lambda \over 2},{{s+1} \over 2};{3 \over 2},{{s+\lambda} \over 2}+{3 \over 4}; {{4t^2} \over {(1+t^2)^2}}\right) \right], $$
(c) the polynomial factors satisfy the functional equation $p_n(s)=(-1)^{\lfloor n/2 \rfloor} p_n(1-s)$, 
and (d) (location of zeros) the polynomial factors of $M_n^\lambda(s)$ have zeros only on the critical line.

{\bf Proposition 6}.  The Mellin transform (1.3) may be written as
$$M_n^\lambda(s)={{\Gamma\left({\lambda \over 2}+{1 \over 4}\right)\Gamma\left({{s+n} \over 2}\right)}\over {2n!\Gamma \left({{s+n+\lambda} \over 2}+{1 \over 4}\right)}}
~_3F_2\left({\lambda \over 2}+{1 \over 4},{{1-n} \over 2},-{n \over 2};{1 \over 2}+\lambda,
1-{{(n+s)} \over 2};1\right).$$
It follows that its polynomial factors satisfy the functional equation $p_n(s)=(-1)^{\lfloor n/2 \rfloor} p_n(1-s)$ and their zeros all lie on the critical line. 

Our results also offer the prospect of developing new representations of the zeta function.

The following section of the paper contains the proof of these Propositions.  Section
3 contains various supporting and reference Lemmas.  Some of these Lemmas present 
results of special function theory that may be of interest in themselves. 
In section 4 we briefly consider certain Mellin transforms of Chebyshev functions of the
first kind.  Finally, we mention some points for further investigation and provide some
discussion of the results.

Within the Askey hierarchy of hypergeometric polynomials, the continuous Hahn polynomials 
(e.g., \cite{andrews} p. 331, \cite{askey}) occur at the $_3F_2(1)$ level.  The connection of the polynomial factors of our Mellin transforms with these polynomials is described after the following section.

\medskip
\centerline{\bf Proof of Propositions}
\medskip

{\it Proposition 1}.  According to Lemma 6 or 8,
the Mellin transforms are of the form 
$$M_n(s)=\Gamma\left({3 \over 4}\right) {{p_n(s) \Gamma\left({{s+\varepsilon} \over 2} \right)} \over {\Gamma\left({s \over 2}+{{2n+3} \over 4}\right)}}, \eqno(2.1)$$
where $\varepsilon=0$ for $n$ even and $=1$ for $n$ odd.  
The recursions (1.2) follow by inserting the form (2.1) into (3.1), and using
the functional equation of the Gamma function, $\Gamma(z+1)=z\Gamma(z)$.  \qed

{\it Proposition 2}.
From Lemma 8(a), up to $s$-{\it independent} factors, the polynomials $p_n$ may be taken as
$$p_n(s)=(n+1){{\Gamma\left({3 \over 4}\right)} \over 2}{{\Gamma\left({{n+s} \over 2}\right)} \over {\Gamma\left({{s+\varepsilon} \over 2}\right)}} ~_3F_2\left({3 \over 4},{{1-n} \over 2},-{n \over 2};{3 \over 2},1-{{(n+s)} \over 2};1\right), \eqno(2.2)$$
wherein $\varepsilon=0$ for $n$ even and $=1$ for $n$ odd.  From the form of the
numerator parameters, the degree of $p_n$ is evident.

By a `Beta transformation' \cite{grad} (p. 850) we have the following integral
representation:
$$~_3F_2\left({3 \over 4},{{1-n} \over 2},-{n \over 2};{3 \over 2},1-{{(n+s)} \over 2};1\right)$$
$$={\sqrt{\pi} \over {2\Gamma^2(3/4)}}\int_0^1 (1-x)^{-1/4}x^{-1/4} ~_2F_1\left({{1-n} \over 2},-{n \over 2};1-{{(n+s)} \over 2};x\right)dx.$$
We use (2.2), together with an $x \to 1-x$ transformation of the $_2F_1$ function
\cite{grad} (p. 1043).  Owing to the poles of the $\Gamma$ function, and that $n$ is a
nonnegative integer, the $_2F_1(x)$ function then transforms to a single $_2F_1(1-x)$ function, and there results
$$p_n(s)={{(n+1)} \over 4} {{\Gamma\left({{n+s} \over 2}\right)} \over {\Gamma\left({{s+\varepsilon} \over 2}\right)}}{\sqrt{\pi} \over {\Gamma(3/4)}}
{{\Gamma\left(1-{{(n+s)} \over 2}\right)} \over {\Gamma\left(1-{s \over 2}\right)}}
{{\Gamma\left({{1+n-s} \over 2}\right)} \over {\Gamma\left({{1-s} \over 2}\right)}}$$
$$ \times \int_0^1 (1-x)^{-1/4}x^{-1/4} ~_2F_1\left({{1-n} \over 2},-{n \over 2};{{s-n+1} 
\over 2};1-x\right)dx$$

$$={{(n+1)} \over 4} {\pi \over {\Gamma\left({{s+\varepsilon} \over 2}\right)}}{\sqrt{\pi} \over {\sin\pi\left({{n+s} \over 2}\right)\Gamma(3/4)}}
{1 \over {\Gamma\left(1-{s \over 2}\right)}}
{{\Gamma\left({{1+n-s} \over 2}\right)} \over {\Gamma\left({{1-s} \over 2}\right)}}$$
$$ \times \int_0^1 (1-x)^{-1/4}x^{-1/4} ~_2F_1\left({{1-n} \over 2},-{n \over 2};{{s-n+1} 
\over 2};x\right)dx.$$
The following observations lead to verification of the functional equation.
When $n$ is even, $\varepsilon=0$, 
$$\Gamma\left({s \over 2}\right)\Gamma\left(1-{s \over 2}\right)={\pi \over {\sin\pi(s/2)}},$$
leaving the denominator factor $\Gamma\left({{1-s} \over 2}\right)$.
When $n$ is odd, $\varepsilon=1$, 
$$\Gamma\left({{s+1} \over 2}\right)\Gamma\left({{1-s} \over 2}\right)={\pi \over {\cos\pi(s/2)}},$$
leaving the denominator factor $\Gamma\left(1-{s \over 2}\right)$.
Hence the factor $(-1)^{\lfloor n/2 \rfloor}$ emerges as $\sin (\pi s/2)/\sin [\pi(n+s)/2] =(-1)^{n/2}$ when $n$ is even and as $\cos (\pi s/2)/\sin [\pi(n+s)/2]=(-1)^{(n-1)/2}$ when 
$n$ is odd, and the functional equation of $p_n(s)$ follows.  

Let $q$ be the shifted polynomial $q(s)=p_n(s+1/2)$, so that $p_n(s)=q(s-1/2)$.  
Furthermore, let $r_i$ be the roots of $q$, $q(s)=c \prod_i(s-r_i)$, where $c$ is a constant.
As a special case (of $\lambda=1$) in the difference equation (2.5), it follows that for
$s$ being a root $r_k$ that
$$-(r_k-n)\left(r_k+{1 \over 2}\right)c \prod_i(r_k+2-r_i)=(r_k+n)\left(r_k-{1 \over 2}\right)
c \prod_i(r_k-2-r_i).$$
The equality of the absolute value of both sides of this equation provides a necessary
condition that Re $r_i=0$.  With all zeros of $q$ being pure imaginary, all of those of
$p_n(s)$ lie on the critical line.  Since the polynomial coefficients are real, in fact,
rational numbers, the zeros of $p_n$ aside from $1/2$ occur as complex conjugates.  \qed

{\it Proposition 3}.  The proof follows that of Proposition 2, noting the integral representation
$$\int_0^1 (1-x)^{-\beta}x^{-\beta} ~_2F_1\left({{1-n} \over 2},-{n \over 2};1-{{(n+s)} \over 2};x\right)dx$$
$$ =2^{2\beta-1} {{\sqrt{\pi} \Gamma(1-\beta)} \over {\Gamma(3/2-\beta)}} ~_3F_2\left(1-\beta,{{1-n} \over 2},-{n \over 2};2(1-\beta),1-{{(n+s)} \over 2};1\right),
\eqno(2.3)$$
with $\beta<1$. 
The $_2F_1(x)$ function is again transformed to a $_2F_1(1-x)$ function and the other
steps are very similar to before.  

The location of the zeros follows from Proposition 5, with the identification 
$\lambda =3/2-2\beta$.  \qed

{\it Corollary 1}.  (a) The initial $\beta=0$ reduction of Proposition 3 to $_2F_1$ form follows from the series definition of the $_3F_2$ function with a shift of summation index and the relations $(1)_j/(2)_j=1/(j+1)$ and $(\kappa)_{j-1}=(\kappa-1)_j/(\kappa-1)$.  The second reduction is a consequence of Gauss summation.  
(b) Similarly, with $m$ a positive integer, $(m+1)_j/(2(m+1))_j$ may be reduced and partial
fractions applied to this ratio.  Then with shifts of summation index, the $_3F_2$ function may
be reduced to a series of $_2F_1(1)$ functions.  These in turn may be written in terms of 
ratios of Gamma functions from Gauss summation.  \qed

{\it Proposition 4}. It follows from Lemma 3(b) that
$$M_{2n}(s)=\sum_{r=0}^n (-1)^{n-r} {{n+r} \choose {2r}}2^{2r} M_0(s+2r),$$
and
$$M_{2n+1}(s)=\sum_{r=0}^n (-1)^{n-r} {{n+r+1} \choose {2r+1}}2^{2r+1} M_0(s+2r+1).$$
By iterating the functional equation of the Gamma function, $\Gamma(z+1)=z\Gamma(z)$, we
obtain
$$M_0(s+2r)=2^rM_0(s)\prod_{q=0}^{r-1}{{s+2q} \over {2s+4q+3}},$$
and
$$M_0(s+2r+1)=2^rM_0(s+1)\prod_{q=0}^{r-1}{{s+2q+1} \over {2s+4q+5}}.$$
Then altogether,
$$M_{2n}(s)=M_0(s)\sum_{r=0}^n (-1)^{n-r} {{n+r} \choose {2r}}2^{3r} \prod_{q=0}^{r-1}{{s+2q} \over {2s+4q+3}},$$
and
$$M_{2n+1}(s)=M_0(s+1)\sum_{r=0}^n (-1)^{n-r} {{n+r+1} \choose {2r+1}}2^{3r+1} \prod_{q=0}^{r-1}{{s+2q} \over {2s+4q+5}}.$$
We next convert the binomial coefficients and products to Pochhammer symbols, and then use
the series definition of the $_3F_2$ function to yield the stated expressions.  
We have
$$\prod_{q=0}^{r-1}(s+2q+\alpha)=2^r \left({{s+\alpha} \over 2}\right)_r$$
and
$$\prod_{q=0}^{r-1}(2s+4q+\beta)=4^r \left({{2s+\beta} \over 4}\right)_r.$$
The binomial coefficient
$${{n+r+c} \choose {2r+c}}={{\Gamma(n+r+c+1)} \over {\Gamma(2r+c+1)\Gamma(n-r+1)}}$$
may be transformed using
$$\Gamma(n+r+c+1)=\Gamma(n+c+1)(n+c+1)_r, ~~~~~~{1 \over {\Gamma(n-r+1)}}={{(-1)^r (-n)_r}
\over {\Gamma(n+1)}},$$
and, as a result of the duplication formula for Pochhammer symbols,
$$\Gamma(2r+c+1)=\Gamma(c+1)4^r \left({{c+1} \over 2}\right)_r \left({{c+2} \over 2}\right)_r 
.  \eqno(2.4)$$
There results
$${{n+r+c} \choose {2r+c}}={{(-1)^r \Gamma(n+c+1)(n+c+1)_r (-n)_r} \over {n! \Gamma(c+1)4^r \left({{c+1} \over 2}\right)_r \left({{c+2} \over 2}\right)_r}}.$$
Substituting this coefficient and the various products, and specializing the parameters
$\alpha$, $\beta$, and $c$, gives the Proposition.  \qed

{\it Proposition 5}.  (a) follows from (\cite{andrews}, p. 303 or \cite{grad}, p. 1030)
$$(n+2)C_{n+2}^\lambda(x)=2(\lambda+n+1)xC_{n+1}^\lambda(x)-(2\lambda+n)C_n^\lambda(x),$$
$C_0^\lambda(x)=1$, and $C_1^\lambda(x)=2\lambda x$.
(b) follows from (\cite{andrews}, p. 302 or \cite{grad}, p. 1029)
$$(1-2xt+t^2)^{-\lambda}=\sum_{n=0}^\infty C_n^\lambda(x)t^n.$$ 
(c) The Mellin transforms may be computed explicitly by several means.  In particular, from
$$C_n^\lambda(x)={{(\lambda)_n} \over {n!}} (2x)^n ~_2F_1\left(-{n \over 2},{{1-n} \over 2};
1-n-\lambda;{1 \over x^2}\right),$$
transformation of the $_2F_1$ function to argument $x^2$ and use of \cite{grad} (p. 850), we find that
$$M_n^\lambda(s)={{(\lambda)_n} \over {n!}} 2^{n-1}\sqrt{\pi}(-i)^n \Gamma\left({\lambda \over
2}+{1 \over 4}\right)\Gamma(1-\lambda-n)$$
$$\times\left\{{{-2i\Gamma\left({{s+1} \over 2}\right)} \over {\Gamma\left({1 \over 2}-
\lambda-{n \over 2}\right)}} {{_3F_2\left({{1-n} \over 2},\lambda+{{n+1} \over 2},{{s+1}
\over 2};{3 \over 2},{3 \over 4}+{{\lambda+s} \over 2};1\right)} \over
{\Gamma\left(-{n \over 2}\right)\Gamma\left({{\lambda+s} \over 2}+{3 \over 4}\right)}}\right.$$
$$\left. + {{\Gamma\left({s \over 2}\right)} \over {\Gamma\left({{1-n} \over 2}\right)}} {{_3F_2\left(-{n \over 2},\lambda+{n \over 2},{s \over 2};{1 \over 2},{1 \over 4}+{{\lambda+s} \over 2};1\right)} \over
{\Gamma\left(1-\lambda-{n \over 2}\right)\Gamma\left({{\lambda+s} \over 2}+{1 \over 4}\right)}}\right \}.$$
The second line of the right member provides the transform for $n$ odd and the third
line for $n$ even.  Then transformation of the $_3F_2$ functions and arguments similar to the 
proof of Proposition 1 may be used to show that the degree of the polynomial factors of the transforms is again $\lfloor n/2 \rfloor$, and that they satisfy the functional equation $p_n(s)=(-1)^{\lfloor n/2 \rfloor} p_n(1-s)$.  

(d) To show that the resulting zeros occur only on Re $s=1/2$ we first demonstrate the  
difference equation
$$[6-4(\lambda+2\lambda n+n^2)-16s+8s(s+1)]\left({{s+\epsilon} \over 2}-1\right)
\left({{s+n+\lambda} \over 2}+{1 \over 4}\right)p_n^\lambda(s)$$
$$+[-9+4(n+\lambda)^2-4(s-1)(s+2)]\left({{s+\epsilon} \over 2}\right)\left({{s+\epsilon} \over 2}-1\right)p_n^\lambda(s+2) \eqno(2.5)$$
$$-4(s-1)(s-2)\left({{s+n+\lambda} \over 2}+{1 \over 4}\right)\left({{s+n+\lambda} \over 2}
-{3 \over 4}\right)p_n^\lambda(s-2)=0,$$
wherein $\epsilon=0$ for $n$ even and $=1$ for $n$ odd.
We apply the ordinary differential equation satisfied by Gegenbauer polynomials (e.g.,
\cite{grad}, p. 1031)
$$(x^2-1)y''(x)+(2\lambda+1)xy'(x)-n(2\lambda+n)y(x)=0.$$
If $f(x)\equiv C_n^\lambda(x)/(1-x^2)^{3/4-\lambda/2}$, we then substitute
$C_n^\lambda(x)=(1-x^2)^{3/4-\lambda/2}f(x)$ into this differential equation.
We then find that
$${1 \over 4}(1-x^2)^{-1/4-\lambda/2}\left[(6-4(\lambda+2\lambda n+n^2)+(-9+4(\lambda+n)^2)x^2
)f(x) \right.$$
$$\left. +4(x^2-1)(-4xf'(x)+(1-x^2)f''(x))\right]=0.$$
It follows that the quantity in square brackets is zero.  We multiply it by $x^{s-1}$
and integrate from $x=0$ to $1$, integrating the $f'$ term once by parts, and the $f''$
term twice by parts.  We determine that the Mellin transforms satisfy the following 
difference equation:
$$[6-4(\lambda+2\lambda n+n^2)-16s+8s(s+1)]M_n^\lambda(s)$$
$$+[-9+4(n+\lambda)^2+16(s+2)-4(s+2)(s+3)]M_n^\lambda(s+2)$$
$$-4(s-1)(s-2)M_n^\lambda(s-2)=0.$$
As follows from either part (c) or Proposition 6, the Mellin transforms are of the form
$$M_n^\lambda(s)={{\Gamma\left({\lambda \over 2}+{1 \over 4}\right)\Gamma\left({{s+\epsilon}
\over 2}\right)} \over {2n!\Gamma\left({{s+n+\lambda} \over 2}+{1 \over 4}\right)}}
p_n^\lambda(s).$$
Noting that the factor $\Gamma\left({\lambda \over 2}+{1 \over 4}\right)/(2n!)$ is 
independent of $s$, and repeatedly applying the functional equation $\Gamma(z+1)=z\Gamma(z)$
leads to (2.5).

Using the shifted polynomial $q(s)=p_n^\lambda(s+1/2)$, so that $p_n^\lambda(s)=q(s-1/2)$ is
convenient.  Then with the translation $s \to s+1/2$ (2.5) gives
$$\left[6-4(\lambda+2\lambda n+n^2)-16\left(s+{1 \over 2}\right)+8\left(s+{1 \over 2}\right)
\left(s+{3 \over 2}\right)\right]\left({{s+\epsilon} \over 2}-{3 \over 4}\right)\left({{s+n+\lambda}\over 2}+{1 \over 2}\right)q(s)$$
$$+\left[-9+4(n+\lambda)^2-4\left(s-{1 \over 2}\right)\left(s+{5 \over 2}\right)\right]
\left({{s+\epsilon} \over 2}+{1 \over 4}\right)\left({{s+\epsilon} \over 2}-{3 \over 4}\right)
q(s+2)$$
$$-4\left(s-{1 \over 2}\right)\left(s-{3 \over 2}\right)\left({{s+n+\lambda} \over 2}
+{1 \over 2}\right)\left({{s+n+\lambda} \over 2}-{1 \over 2}\right)q(s-2)=0.$$
After some simplification, in particular cancelling a factor of $s+n+\lambda+1$ on both 
sides, it follows that if $r_k$ is a root of $q$, $q(r_k)=0$, that
$$-(r_k-n-\lambda+1)\left(r_k+\epsilon+{1 \over 2}\right)\left(r_k+\epsilon-{3 \over 2}\right)q(r_k+2)$$
$$=(r_k+n+\lambda-1)\left(r_k-{1 \over 2}\right)\left(r_k-{3 \over 2}\right)q(r_k-2).$$
It may be noted that when $n$ is even, a factor of $r_k-3/2$ cancels on both sides, and
when $n$ is odd, a factor of $r_k-1/2$ cancels on both sides.  In either case,
equality of the absolute value of both sides provides a necessary condition that
Re $r_i=0$ for all the zeros of $q$.  Hence the zeros of $p_n^\lambda(s)$ lie on the
critical line.  \qed

{\it Proposition 6}.  We use the series representation (\cite{rainville}, p. 278 (6))
$$C_n^\lambda(x)=\sum_{k=0}^{[n/2]} {{(2\lambda)_n x^{n-2k} (x^2-1)^k} \over {4^k k! (\lambda+
1/2)_k(n-2k)!}},$$
and recall a form of the Beta integral,
$$\int_0^1 x^{a-1}(1-x^2)^{b-1}dx={1 \over 2}B\left({a \over 2},b\right), ~~~~\mbox{Re} ~a>0,
~~\mbox{Re} ~b>0.$$
Then
$$M_n^\lambda(s)=\sum_{k=0}^{[n/2]} {{(2\lambda)_n (-1)^k} \over {4^k k! (\lambda+
1/2)_k(n-2k)!}}\int_0^1x^{s+n-2k-1}(1-x^2)^{k+\lambda/2-3/4}dx$$
$$=\sum_{k=0}^{[n/2]} {{(2\lambda)_n (-1)^k} \over {4^k k! (\lambda+
1/2)_k(n-2k)!}}{1 \over 2}B\left({{s+n} \over 2}-k,k+{\lambda \over 2}+{1 \over 4}\right)$$
$$={{(2\lambda)_n} \over {2\Gamma\left({{s+n+\lambda} \over 2}+{1 \over 4}\right)}}
\sum_{k=0}^{[n/2]} {{(-1)^k} \over {4^k k!}}{{\Gamma\left({{s+n} \over 2}-k\right)
\Gamma\left(k+{\lambda \over 2}+{1 \over 4}\right)} \over {(\lambda+1/2)_k(n-2k)!}}$$
$$={{\Gamma\left({\lambda \over 2}+{1 \over 4}\right)\Gamma\left({{s+n} \over 2}\right)}\over {2n!\Gamma \left({{s+n+\lambda} \over 2}+{1 \over 4}\right)}}
~_3F_2\left({\lambda \over 2}+{1 \over 4},{{1-n} \over 2},-{n \over 2};{1 \over 2}+\lambda,
1-{{(n+s)} \over 2};1\right).$$

The above has used the duplication formula for the Gamma function so that
$${1 \over {(n-2k)!}}={1 \over {\Gamma(n-2k+1)}}={4^k \over {n!}}\left(-{n \over 2}\right)_k
\left({{1-n} \over 2}\right)_k.$$
The latter form of $M_n^\lambda(s)$ exhibits the key feature of a denominator parameter twice a numerator parameter and we identify $\beta=3/4-\lambda/2$ in Proposition 3.  The stated 
properties follow in accord with that result.  \qed

{\it Remarks}.  With a simple change of variable, the Mellin transforms of Proposition 5 may
be gotten from \cite{grad} (p. 830).  Alternative forms of these transforms and their
generating functions may be obtained by using various expressions from \cite{rainville} 
(pp. 279--280).  

\medskip
\centerline{\bf Connection with continuous Hahn polynomials}
\medskip

These polynomials are given by (e.g., \cite{andrews} p. 331, \cite{askey})
$$p_n(x;a,b,c,d)=i^n {{(a+c)_n(a+d)_n} \over {n!}} ~_3F_2\left(-n,n+a+b+c+d-1,a+ix;a+c,a+d;1
\right).$$
By using the second transformation of a terminating $_3F_2(1)$ series given in the Appendix,
we have
$$p_n(x;a,b,c,d)={i^n \over {n!}}(a+b+c+d+n-1)_n(1-b-n-ix)_n $$
$$ \times ~_3F_2(1-b-c-n,1-b-d-n,-n;2-a-b-c-d-2n,1-b-n-ix;1).$$
Then comparing with the $_3F_2(1)$ function of Proposition 6, we see that our polynomial
factors are proportional to the continuous Hahn polynomial
$$p_n\left(-{{is} \over 2};{1 \over 4}-{\lambda \over 2}-n,0,{3 \over 4}-{\lambda \over 2}-n,
{1 \over 2}\right).$$
The continuous Hahn polynomials are orthogonal on the line with respect to the measure
$${1 \over {2\pi}}\Gamma(a+ix)\Gamma(b+ix)\Gamma(c-ix)\Gamma(d-ix)dx.$$
Due to the Parseval relation for the Mellin transform, 
$$\int_0^\infty f(x)g^*(x)dx={1 \over {2\pi i}}\int_{(0)} ({\cal M}f)(s) ({\cal M}g)^*(s)ds,$$
the polynomial factors $p_n(1/2+it)$ form an orthogonal family with respect to a suitable
measure with $\Gamma$ factors.  Since orthogonal polynomials have real zeros, this approach
provides another way of showing that $p_n(s)$ has zeros only on the critical line.

\medskip
\centerline{\bf Lemmas}
\medskip

{\bf Lemma 1}. (a)
$$M_{mn-1}(s)=\int_0^1 {x^{s-1} \over {(1-x^2)^{1/4}}} U_{m-1}[T_n(x)]U_{n-1}(x)dx$$
$$=\int_0^1 {x^{s-1} \over {(1-x^2)^{1/4}}} U_{n-1}[T_m(x)]U_{m-1}(x)dx,$$
and (b)
$${1 \over 2}[M_{m+n-1}(s)+M_{m-n-1}(s)]=\int_0^1 x^{s-1} T_n(x)U_{m-1}(x) {{dx} \over
{(1-x^2)^{1/4}}}.$$
For polynomials of negative index, we have $U_{-n}(x)=-U_{n-2}(x)$.

{\it Proof}.  Part (a) follows from the composition-product property
$$U_{mn-1}(x)=U_{m-1}[T_n(x)]U_{n-1}(x)=U_{n-1}[T_m(x)]U_{m-1}(x),$$
and (b) from $(1/2)[U_{m+n-1}(s)+U_{m-n-1}(s)]=T_n(x)U_{m-1}(x)$.

{\bf Lemma 2}.  (Mixed recursion relation.)
$$M_n(s)=2M_{n-1}(s+1)-M_{n-2}(s),  \eqno(3.1)$$
and 
$$M_0(s)={{\Gamma\left({3 \over 4}\right)} \over 2}{{\Gamma\left({s \over 2}\right)} \over
{\Gamma\left({s \over 2}+{3 \over 4}\right)}}, ~~~~~~
M_1(s)=\Gamma\left({3 \over 4}\right){{\Gamma\left({{s+1} \over 2}\right)} \over
{\Gamma\left({s \over 2}+{5 \over 4}\right)}}.  \eqno(3.2)$$
Here $M_1(s)=2M_0(s+1)$ and $M_0(s)=B(s/2,3/4)/2$. 

{\it Proof}.  The recursion (3.1) follows from
$U_{n+1}(x)-2xU_n(x)+U_{n-1}(x)=0$.  With $U_0=1$, with a simple change of variable we have
$$M_0(s)={1 \over 2}\int_0^1 u^{s/2-1}(1-u)^{-1/4}du={1 \over 2}B\left({s\over 2},{3 \over 4}
\right),$$
giving (3.2).  Since $U_1(x)=2x$, $M_1(s)=2M_0(s+1)$ immediately follows. \qed

Here is a second proof of the Lemma.  By using the definition (3.11) we have the
trigonometric form
$$M_n(s)=\int_0^{\pi/2} \cos^{s-1} \theta (\sin^{-1/2} \theta) \sin[(n+1)\theta]d\theta.$$
Then the recursion (3.1) follows from the identity
$\sin(n+1)\theta+\sin(n-1)\theta=2\cos\theta \sin n\theta$.  \qed

{\bf Lemma 3}. (a)
$$M_n(s+m)={1 \over 2^m}\sum_{r=0}^m {m \choose r} M_{m+n-2r}(s)$$
and (b)
$$M_n(s)=\sum_{r=0}^k (-1)^{k-r} {k \choose r} 2^r M_{n-2k+r}(s+r).$$

{\it Proof}.  (a) For $m<n$, we have
$$x^mU_n(x)={1 \over 2^m}\sum_{r=0}^m {m \choose r} U_{m+n-2r}(x),$$
provable by induction, using the property $xU_n(x)={1 \over 2}[U_{n+1}(x)+U_{n-1}(x)].$  
Then by the definition (1.1) the result follows.  
For the inductive step, we easily have
$$x^{m+1}U_n(x)={1 \over 2^m}\sum_{r=0}^m {m \choose r} xU_{m+n-2r}(x)$$
$$={1 \over 2^{m+1}}\sum_{r=0}^m {m \choose r} \left[U_{m+n-2r+1}(x)+U_{m+n-2r-1}(x)\right]$$
$$={1 \over 2^{m+1}}\sum_{r=0}^{m+1} \left[{m \choose r} + {m \choose {r-1}}\right]
U_{m+n-2r+1}$$
$$={1 \over 2^{m+1}}\sum_{r=0}^{m+1} {{m+1} \choose r}U_{m+n-2r+1}.$$
\qed

(b) follows from the binomial transform
$$f_n=\sum_{i=0}^n {n \choose i}g_i \iff g_n=\sum_{i=0}^n (-1)^{n+i} {n \choose i}f_i,$$
which is based upon the orthogonality property of binomial coefficients
$$\sum_{k=m}^n (-1)^{k+m} {n \choose k}{k \choose m}=\delta_{nm},$$
where $\delta_{nm}$ is the Kronecker symbol.
\qed

We have the generating function
$$\sum_{k=0}^\infty U_k(x) t^k ={1 \over {1-2tx+t^2}}.  \eqno(3.3)$$
{\bf Lemma 4}.  (Generating function of Mellin transforms.)  For Re $s>0$,
$$G(t,s) \equiv \sum_{k=0}^\infty M_k(s)t^k =\int_0^1 {1 \over {(1-x^2)^{1/4}}}{x^{s-1} \over 
{(1-2tx+t^2)}}dx$$
$$={1 \over {(1+t^2)}}{{\Gamma\left({3 \over 4}\right)} \over 2}\left[{{\Gamma\left({s \over 2}\right)} \over {\Gamma\left({s \over 2}+{3 \over 4}\right)}} ~_2F_1\left(1,{s \over 2};
{{2s+3} \over 4};{{4t^2} \over {(1+t^2)^2}}\right) \right.$$
$$\left. + {{2t} \over {(1+t^2)}}{{\Gamma\left({{s+1} \over 2}\right)} \over {\Gamma\left({s \over 2}+{5 \over 4}\right)}} ~_2F_1\left(1,{{s+1} \over 2};{{2s+5} \over 4}; {{4t^2} \over {(1+t^2)^2}}\right) \right].  \eqno(3.4)$$
As given below, the first line of the right member of (3.4) furnishes $M_{2k}(s)$ and the 
second line, $M_{2k+1}(s)$.

{\it Proof}.  Using (3.3), we first binomially expand so that
$$\int_0^1 {1 \over {(1-x^2)^{1/4}}}{x^{s-1} \over {(1-2tx+t^2)}}dx
=\sum_{\ell=0}^\infty {{-{1 \over 4}} \choose \ell}(-1)^\ell \int_0^1 {x^{2\ell+s-1} \over
{1-2tx +t^2}}dx$$
$$={1 \over {1+t^2}}\sum_{\ell=0}^\infty {{-{1 \over 4}} \choose \ell}{{(-1)^\ell } \over {(2\ell+s)}} ~_2F_1\left(1,2\ell+s;1+2\ell+s;{{2t} \over {1+t^2}}\right).$$
Then we interchange sums, and separate terms of even and odd summation index:
$$\sum_{k=0}^\infty M_k(s)t^k = {1 \over {1+t^2}}\sum_{\ell=0}^\infty {{-{1 \over 4}} \choose \ell}{{(-1)^\ell } \over {(2\ell+s)}} \sum_{j=0}^\infty {{(2\ell+s)_j} \over {(2\ell+s+1)_j}}
\left({{2t} \over {1+t^2}}\right)^j$$
$$={{\Gamma(3/4)} \over {2(1+t^2)}} \sum_{j=0}^\infty {{\Gamma\left({{j+s} \over 2}\right)} \over {\Gamma\left({3 \over 4}+{{j+s} \over 2}\right)}} \left({{2t} \over {1+t^2}}\right)^j$$
$$={{\Gamma(3/4)} \over {2(1+t^2)}} \left[\sum_{m=0}^\infty {{\Gamma\left(m+{s \over 2}\right)} \over {\Gamma\left({3 \over 4}+{s \over 2}+m\right)}} \left({{2t} \over {1+t^2}}\right)^{2m}
+\sum_{m=0}^\infty {{\Gamma\left(m+1+{s \over 2}\right)} \over {\Gamma\left({5 \over 4}+{s \over 2}+m\right)}} \left({{2t} \over {1+t^2}}\right)^{2m+1}
\right]$$
$$={{\Gamma(3/4)} \over {2(1+t^2)}} \left[{{\Gamma\left({s \over 2}\right)} \over {\Gamma\left({s \over 2}+{3 \over 4}\right)}}\sum_{m=0}^\infty {{(1)_m} \over {m!}}
{{\left({s \over 2}\right)_m} \over {\left({3 \over 4}+{s \over 2}\right)_m}} \left({{2t} \over {1+t^2}}\right)^{2m}\right.$$
$$\left. +{{\Gamma\left({s \over 2}+1\right)} \over {\Gamma\left({s \over 2}+{5 \over 4}\right)}}\sum_{m=0}^\infty {{(1)_m} \over {m!}}{{\left(1+{s \over 2}\right)_m} \over {\left({5 \over 4}+{s \over 2}\right)_m}} \left({{2t} \over {1+t^2}}\right)^{2m+1} \right].$$
\qed

We mention a second method of proof of Lemma 4.  With the use of partial fractions, we have
$$\sum_{k=0}^\infty M_k(s)t^k =\int_0^1 {{(1-x^2)^{3/4}} \over {(1-x^2)}}{x^{s-1} \over 
{(1-2tx+t^2)}}dx$$
$$=\int_0^1 x^{s-1}(1-x^2)^{3/4}\left\{{1 \over 2}\left[{1\over {(1+t)^2}}{1 \over {(1+x)}}
+ {1\over {(1-t)^2}}{1 \over {(1-x)}}\right]-{{4t^2} \over {(1-t^2)^2(1-2tx+t^2)}}\right\}dx.$$
The Beta function suffices to evaluate the first two integrals on the right side, while the
$_2F_1$ function is required for the third,
$$\sum_{k=0}^\infty M_k(s)t^k = {{\Gamma\left({3 \over 4}\right)} \over 2}{1 \over {(1-t^2)^2}}
\left[(1+t^2){{\Gamma\left({s \over 2}\right)} \over {\Gamma\left({3 \over 4}+{s \over 2}\right)
}} + 2t{{\Gamma\left({{s+1} \over 2}\right)} \over {\Gamma\left ({5 \over 4}+{s \over 2}\right)
}} \right]$$
$$-{{4t^2} \over {(1-t^2)^2}}{{\Gamma\left({7 \over 4}\right)} \over {2(1+t^2)}} \left[
{{\Gamma\left({s \over 2}\right)} \over {\Gamma\left({7 \over 4}+{s \over 2}\right)
}} ~_2F_1\left(1,{s \over 2};{7 \over 4}+{s \over 2};{{4t^2} \over {(1+t^2)^2}}\right)\right.$$
$$\left.
+ {{2t} \over {(1+t^2)}}{{\Gamma\left({{s+1} \over 2}\right)} \over {\Gamma\left({9 \over 4}
+{s \over 2}\right)}} ~_2F_1\left(1,{{s+1} \over 2};{9 \over 4}+{s \over 2};{{4t^2} \over {(1+t^2)^2}}\right) \right].$$
Then contiguous relations for the $_2F_1$ function (e.g., \cite{grad}, pp. 1044-1045) may be
used to show the equivalence with (3.4). \qed

{\bf Corollary 2}.  (Summation identity.)  Let $t$ be sufficiently small.  Then
$${1 \over {(1+t^2)}}{{\Gamma\left({3 \over 4}\right)} \over 2}\left[{{\Gamma\left({s \over 2}\right)} \over {\Gamma\left({s \over 2}+{3 \over 4}\right)}} ~_2F_1\left(1,{s \over 2};
{{2s+3} \over 4};{{4t^2} \over {(1+t^2)^2}}\right) \right.$$
$$\left. + {{2t} \over {(1+t^2)}}{{\Gamma\left({{s+1} \over 2}\right)} \over {\Gamma\left({s \over 2}+{5 \over 4}\right)}} ~_2F_1\left(1,{{s+1} \over 2};{{2s+5} \over 4}; {{4t^2} \over {(1+t^2)^2}}\right) \right]$$
$$={{\Gamma\left({3 \over 4}\right)} \over 2}\sum_{k=0}^\infty (-1)^k t^{2k}\left[{{\Gamma\left({s \over 2}\right)} \over {\Gamma\left({s \over 2}+{3 \over 4}\right)}} ~_3F_2\left({{1-k} \over 2},{s \over 2},-{k \over 2};{1 \over 2},
{{2s+3} \over 4};{4 \over t^2}\right) \right.$$
$$\left. -{{2k\Gamma\left({{s+1} \over 2}\right)} \over {t\Gamma\left({s \over 2}+{5 \over 4}\right)}} ~_3F_2\left({{1-k} \over 2},1-{k \over 2},{{s+1} \over 2};{3 \over 2},{{2s+5} \over 4}; {4 \over t^2}\right) \right].$$

{\it Proof}.  This follows by equating the result for $G(t,s)$ of Lemma 4 with the 
geometric series expansion and integration
$$G(t,s)=\int_0^1 {x^{s-1} \over {(1-x^2)^{1/4}}} \sum_{k=0}^\infty (-1)^k t^k(t-2x)^k dx.$$
\qed


{\it Remarks}.
Transformations of the $_2F_1$ functions in Lemma 4 include the following.
$$~_2F_1\left(1,{s \over 2};{{2s+3} \over 4};{{4t^2} \over {(1+t^2)^2}}\right)
=\left({{1+t^2} \over {1-t^2}}\right)^2 ~_2F_1\left(1,{3 \over 4};{{2s+3} \over 4};-{{4t^2}
\over {(1-t^2)^2}}\right),$$
and
$$~_2F_1\left(1,{{s+1} \over 2};{{2s+5} \over 4}; {{4t^2} \over {(1+t^2)^2}}\right) 
=\left({{1+t^2} \over {1-t^2}}\right)^2 ~_2F_1\left(1,{3 \over 4};{{2s+5} \over 4};-{{4t^2}
\over {(1-t^2)^2}}\right).$$
 
Letting $'$ denote differentiation with respect to $t$, (3.4) is consistent with the
differential equation
$$G'(t,s)+{1 \over {2t}}G(t,s)={{(t^2-1)} \over {2t}}\int_0^1 {x^{s-1} \over {(1-x^2)^{1/4}}}
{1 \over {(1-2xt+t^2)^2}}dx,$$
obtained by using partial fractions.

{\bf Lemma 5}.  Expansion of certain Gauss hypergeometric functions.  (a)  The coefficient
of $t^{2m}$ in the function
$${1 \over {(1+t^2)}} ~_2F_1\left(a,b;c;{{4t^2} \over {(1+t^2)^2}}\right)$$
is given by
$${{4^m \Gamma(c)\Gamma(a+m)\Gamma(b+m)} \over {\Gamma(a)\Gamma(b)m!\Gamma(c+m)}}
~_4F_3\left({1 \over 2}-m,1-c-m,-m,-m;1-a-m,1-b-m,-2m;1\right).  \eqno(3.5)$$
(b) The coefficient
of $t^{2m}$ in the function
$${1 \over {(1+t^2)^2}} ~_2F_1\left(a,b;c;{{4t^2} \over {(1+t^2)^2}}\right)$$
is given by
$${{4^m \Gamma(c)\Gamma(a+m)\Gamma(b+m)} \over {\Gamma(a)\Gamma(b)m!\Gamma(c+m)}}
~_4F_3\left(-{1 \over 2}-m,1-c-m,-m,-m;1-a-m,1-b-m,-1-2m;1\right).  \eqno(3.6)$$  

{\it Proof}.  We show part (a), and (b) follows similarly.  The method is to binomially
expand, and then to interchange sums.  We have
$${1 \over {(1+t^2)}} ~_2F_1\left(a,b;c;{{4t^2} \over {(1+t^2)^2}}\right)
=\sum_{j=0}^\infty \sum_{\ell=0}^\infty {{(a)_j(b)_j} \over {(c)_j}} {4^j \over {j!}}
{{-2j-1} \choose \ell} t^{2(j+\ell)}$$
$$=\sum_{m=0}^\infty \sum_{\ell=0}^m {{(a)_{m-\ell}(b)_{m-\ell}} \over {(c)_{m-\ell}}} {4^{m-\ell} \over {j!}} {{-2(m-\ell)-1} \choose \ell} t^{2m}.$$
The final transformation to hypergeometric form is accomplished through the relations
$$(\lambda)_{n-k}={{(-1)^k (\lambda)_n} \over {(1-\lambda-n)_k}}, ~~~~~~
(-2m)_{2\ell}=4^\ell (-m)_\ell\left({1 \over 2}-m\right)_\ell,$$
and
$${{-2(m-\ell)-1} \choose \ell}={{\Gamma(2\ell-2m)} \over {\ell! \Gamma(\ell-2m)}}
={1 \over {\ell!}}{{(-2m)_{2\ell}} \over {(-m)_{2\ell}}}={1 \over {\ell!}}{{4^\ell (-m)_\ell\left({1 \over 2}-m\right)_\ell} \over {(-m)_{2\ell}}}.$$
\qed

There are two evident reductions from Lemma 5.  These occur for $a=1$ in both parts (a)
and (b), and for $a=1/2$ in (a) and for $a=3/2$ in (b).  The case of $a=1$ is of specific
interest here, and is next recorded.  
\newline{\bf Corollary 3}.  (a) The coefficient of $t^{2m}$ in the function
$${1 \over {(1+t^2)}} ~_2F_1\left(1,b;c;{{4t^2} \over {(1+t^2)^2}}\right)$$
is  
$${{4^m \Gamma(c)\Gamma(b+m)} \over {\Gamma(b)\Gamma(c+m)}}
~_3F_2\left({1 \over 2}-m,1-c-m,-m,;1-b-m,-2m;1\right).  \eqno(3.7)$$
(b) The coefficient of $t^{2m}$ in the function
$${1 \over {(1+t^2)^2}} ~_2F_1\left(1,b;c;{{4t^2} \over {(1+t^2)^2}}\right)$$
is  
$${{4^m \Gamma(c)\Gamma(b+m)} \over {\Gamma(b)\Gamma(c+m)}}
~_3F_2\left(-{1 \over 2}-m,1-c-m,-m,;1-b-m,-1-2m;1\right).  \eqno(3.8)$$

The following comes from the application of Corollary 2 to the generating function (3.4).
{\newline \bf Lemma 6}.  Let $k \geq 0$.  Then
$$M_{2k}(s)={{\Gamma\left({3 \over 4}\right)} \over 2}{{4^k \Gamma\left(k+{s \over 2}\right)} \over {\Gamma\left(k+{{2s+3} \over 4}\right)}} ~_3F_2\left({1 \over 2}-k,-k-{s \over 2}+{1 \over 4},-k;1-{s \over 2}-k,-2k;1\right), \eqno(3.9a)$$
and 
$$M_{2k+1}(s)=\Gamma\left({3 \over 4}\right){{4^k \Gamma\left(k+{{s+1} \over 2}\right)} \over {\Gamma\left(k+{{2s+5} \over 4}\right)}} ~_3F_2\left(-{1 \over 2}-k,-k-{s \over 2}-{1 \over 4},-k;{{1-s} \over 2}-k,-1-2k;1\right). \eqno(3.9b)$$

{\it Remarks}.
From (3.9) expressions for the special values $M_{2k+1}(0)$ and $M_{2k+1}(1)$ and
$M_{2k}(1)$ may be written.

These transforms may be re-expressed by using Thomae's identity \cite{andrews} (p. 143)
$$_3F_2(a,b,c;d,e;1)={{\Gamma(d)\Gamma(e)\Gamma(w)} \over {\Gamma(a)\Gamma(w+b)\Gamma(w+c)}}
~_3F_2(d-a,e-a,w;w+b,w+c;1),$$
where $w=d+e-a-b-c$.

The following allows us to obtain other equivalent hypergeometric forms for the Mellin
transforms, and hence for the polynomial factors.
{\newline \bf Lemma 7}.  (a)
$$U_n(x)=\sum_{k=0}^{[n/2]} {{n+1} \choose {2k+1}} (x^2-1)^k x^{n-2k}
= (n+1)x^n ~_2F_1\left({{1-n} \over 2},-{n \over 2};{3 \over 2};1-{1 \over x^2}\right),$$
(b)
$$U_n(x)=\sum_{k=0}^{[n/2]} (-1)^k{{n-k} \choose k} (2x)^{n-2k}
= (2x)^{2n} ~_2F_1\left({{1-n} \over 2},-{n \over 2};-n;{1 \over x^2}\right),$$
and (c)
$$T_n(x)=\sum_{k=0}^{[n/2]} {n \choose {2k}} (x^2-1)^k x^{n-2k}
= x^n ~_2F_1\left({{1-n} \over 2},-{n \over 2};{1 \over 2};1-{1 \over x^2}\right).$$

{\it Proof}.  These expressions follow by applying quadratic transformations (e.g., 
\cite{grad}, p. 1043) to known $_2F_1$ forms of $U_n$ and $T_n$,
$$U_n(x) = (n+1) ~_2F_1\left(n+2,-n;{3 \over 2};{{1-x} \over 2}\right), \eqno(3.10)$$
and
$$T_n(x)= ~_2F_1\left(n,-n;{1 \over 2};{{1-x} \over 2}\right).$$

{\bf Lemma 8}. (a)
$$M_n(s)=(n+1){{\Gamma\left({3 \over 4}\right)} \over 2}{{\Gamma\left({{n+s} \over 2}\right)} \over {\Gamma\left({{n+s} \over 2}+{3 \over 4}\right)}} ~_3F_2\left({3 \over 4},{{1-n} \over 2},-{n \over 2};{3 \over 2},1-{{n+s} \over 2};1\right),$$
and (b)
$$M_n(s)=2^{n-1}{{\Gamma\left({3 \over 4}\right)\Gamma\left({{n+s} \over 2}\right)} \over {\Gamma\left({{n+s} \over 2}+{3 \over 4}\right)}} ~_3F_2\left({{1-n} \over 2},-{n \over 2},{1 \over 4}-{{(n+s)} \over 2};-n,1-{{n+s} \over 2};1\right).$$

{\it Proof}.  The interchange of finite summation and the integration of (1.1) is used in
both instances.  \qed

{\bf Lemma 9}. (Other generating functions for $U_n(x)$).  Let $y=\cos^{-1} x$,
$I_0$ be the modified Bessel function of the first kind, $t \neq 0$, and $j\geq 1$ be an integer.  Then (a)
$$\sum_{n=0}^\infty {{U_n(x)} \over {(n+1)!}}t^n ={e^{xt} \over t} {{\sin(t\sqrt{1-x^2})} \over
\sqrt{1-x^2}}, $$
(b)
$$\sum_{n=0}^\infty {{U_n(x)} \over {((n+1)!)^2}}t^n = {i \over {2t}}{1 \over \sqrt{1-x^2}}
\left[I_0\left(2e^{-iy/2}\sqrt{t}\right)-I_0\left(2e^{iy/2}\sqrt{t}\right)\right],$$
and (c)
$$\sum_{n=0}^\infty {{U_n(x)} \over {((n+1)!)^j}}t^n = {i \over {2t\sqrt{1-x^2}}}\left[
~_0F_{j-1}(\cdot;1,1,\ldots,1;e^{-iy}t)-_0F_{j-1}(\cdot;1,1,\ldots,1;e^{iy}t)\right].$$
\noindent
As $U_n(\pm 1)=(\pm 1)^n (n+1)$, (a) includes the cases $x \to \pm 1$.

{\it Proof}.  These relations are based upon the definition
$$U_n(x)={{\sin[(n+1)\cos^{-1}x]} \over \sqrt{1-x^2}}. \eqno(3.11)$$
For part (a) we use this definition along with the relation
$$e^{te^{-iy}}-e^{te^{-iy}}=e^{t\cos y}[e^{-it\sin y}-e^{it\sin y}]=e^{tx}(-2i)\sin(t\sin y).$$
(b) is a special case of (c) as $_0F_1(1;x)=I_0(2\sqrt{x})$.
For part (c), 
$$\sum_{n=0}^\infty {{U_n(x)} \over {((n+1)!)^j}}t^n ={1 \over \sqrt{1-x^2}}\sum_{n=0}^\infty
{{\sin[(n+1)y] t^n} \over {((n+1)!)^j}}={1 \over \sqrt{1-x^2}}\sum_{n=0}^\infty
{{\sin (ny) t^{n-1}} \over {n!(1)_n\cdots(1)_n}},$$
and this implies the result.  \qed


{\bf Lemma 10}. (`Beta transformation' of Chebyshev polynomial).  Let Re $r>-1$ and Re $q>-1$.
Then
$$\int_0^1 x^r (1-x)^q U_n(x)dx=(n+1)B(r+1,q+1) ~_3F_2\left(n+2,-n,q+1;{3 \over 2},q+r+2;{1 \over 2}\right). \eqno(3.12)$$

{\it Proof}.  This result follows from (3.10) and writing
$$\int_0^1 x^r (1-x)^q U_n(x)dx=(n+1) \int_0^1 x^r (1-x)^q~_2F_1\left(n+2,-n;{3 \over 2};{{1-x} \over 2}\right)dx$$
$$=(n+1) \int_0^1 (1-y)^r y^q~_2F_1\left(n+2,-n;{3 \over 2};{y \over 2}\right)dy.$$
\qed

Even the $q=0$ case of (3.12) is not commonly given in integral tables.  This case may be
used to alternatively develop the Mellin transforms (1.1), as we next briefly indicate.

{\bf Lemma 11}.
$$M_n(s)=(n+1)\sum_{j=0}^n {{(n+2)_j(-n)_j} \over {(3/2)_j 2^js(s+1)_j}}~_3F_2\left({1 \over 4},
{{s+1} \over 2},{s \over 2};{{s+j+1} \over 2},{{s+j} \over 2}+1;1\right).$$

{\it Proof}.  This follows from the series and integral
$$M_n(s)=\sum_{\ell=0}^\infty (-1)^\ell {{-1/4} \choose \ell} \int_0^1 x^{s+2\ell-1} U_n(x)dx,$$
applying (3.12), and then interchanging sums.  For the sum over $\ell$, (2.4) is used,
$$\sum_{\ell=0}^\infty (-1)^\ell {{-1/4} \choose \ell} {1 \over {(s+2\ell)}}{1 \over {(s+2\ell+1)_j}}= \sum_{\ell=0}^\infty {{(1/4)_\ell} \over {\ell !}}{1 \over s}{{(s/2)_\ell}
\over {\left({s \over 2}+1\right)_\ell}}{1 \over {(s+2\ell+1)_j}}$$
$$= \sum_{\ell=0}^\infty {{(1/4)_\ell} \over {\ell !}}{1 \over s}{{(s/2)_\ell}
\over {\left({s \over 2}+1\right)_\ell}}{{\Gamma(s+1)} \over {\Gamma(s+j+1)}}
{{\left({{s+1} \over 2}\right)_\ell} \over {\left({{s+j+1} \over 2}\right)_\ell}}
{{\left({s \over 2}+1\right)_\ell} \over {\left({{s+j} \over 2}+1\right)_\ell}}$$
$$= \sum_{\ell=0}^\infty {{(1/4)_\ell} \over {\ell !s(s+1)_j}}
{{\left({s \over 2}\right)_\ell} \over {\left({{s+j+1} \over 2}\right)_\ell}}
{{\left({{s+1} \over 2}\right)_\ell} \over {\left({{s+j} \over 2}+1\right)_\ell}}$$
$$={1 \over {s(s+1)_j}}~_3F_2\left({1 \over 4},{{s+1} \over 2},{s \over 2};{{s+j+1} \over 2},{{s+j} \over 2}+1;1\right).$$
\qed

{\bf Lemma 12}. (a) (A connection of the Chebyshev polynomial of the second kind with the Legendre polynomial $P_n$). 
$$U_m(x)=\sum_{k=0}^m P_k(x)P_{m-k}(x),$$
(b)
$${1 \over 2}U_{m+1}'(x)=\sum_{k=0}^m U_k(x)U_{m-k}(x)={1 \over {2(x^2-1)}}[(m+1)xU_{m+1}(x)
-(m+2)U_m(x)],$$
and (c)
$$C_m^{\lambda_1+\lambda_2}(x)=\sum_{k=0}^m C_k^{\lambda_1}(x)C_{m-k}^{\lambda_2}(x).$$

{\it Proof}.  (a) The generating function (3.3) is used, along with
$$\sum_{k=0}^\infty P_k(x) t^k ={1 \over \sqrt{1-2tx+t^2}}.$$
Since
$$\sum_{k=0}^\infty U_k(x) t^k ={1 \over {1-2tx+t^2}}=\left(\sum_{k=0}^\infty P_k(x) t^k\right)^2$$
holds over a common range of $t$, the result follows.  

(b) The generating function 
$$\sum_{k=0}^\infty C_k^\lambda(x) t^k ={1 \over {(1-2tx+t^2)^\lambda}}$$
is used, together with (3.3) and the expression for $C_n^2(x)$ given in (5.2).  
(c) Again uses the generating function given just above.  \qed

{\bf Lemma 13}. (Limit relation).  
$$\lim_{\lambda \to \infty} {{C_n^\lambda(x)} \over {C_n^\lambda(1)}} =x^n.$$

{\it Proof}.  
We have
$$\lim_{\lambda \to \infty} {{C_n^\lambda(x)} \over {C_n^\lambda(1)}} = \lim_{\lambda \to \infty} ~_2F_1\left(2\lambda+n,-n;\lambda+{1 \over 2};{{1-x} \over 2}\right),$$
and, by Stirling's formula,
$${{(2\lambda+n)_k} \over {\left(\lambda+{1 \over 2}\right)_k}}=2^k\left[1+O\left({1 \over \lambda}
\right)\right].$$
Then
$$\lim_{\lambda \to \infty} {{C_n^\lambda(x)} \over {C_n^\lambda(1)}} =
\sum_{k=0}^n {{(-n)_k} \over {k!}}(1-x)^k=\sum_{k=0}^n (-1)^k{n \choose k}(1-x)^k=x^n.$$
\qed

{\bf Lemma 14}.  (Another generating function for $M_n^\lambda(s)$).  Let $J_\nu$ be the 
Bessel function of the first kind.  Then for $\lambda >-1/2$,
$$\sum_{n=0}^\infty {{M_n^\lambda(s)} \over {C_n^\lambda(1)}}{t^n \over {n!}}
=2^{\lambda-1/2}\Gamma\left(\lambda+{1 \over 2}\right)\int_0^1 {x^{s-1} \over {(1-x^2)^{3/4-\lambda/2}}} e^{xt}(\sqrt{1-x^2}t)^{1/2-\lambda}J_{\lambda-1/2}(\sqrt{1-x^2}t)dx.$$

{\it Proof}.  From the Laplace-type integral (5.1) follows
$$\sum_{n=0}^\infty{{C_n^\lambda(x)} \over {C_n^\lambda(1)}}{t^n \over {n!}} =2^{\lambda-1/2}\Gamma\left(\lambda+{1 \over 2}\right)e^{xt} (\sqrt{1-x^2}t)^{1/2-\lambda}J_{\lambda-1/2}(\sqrt{1-x^2}t).  \eqno(3.13)$$
Then the definition (1.3) is applied.  \qed

{\bf Lemma 15}.  
$$\left.{{C_n^\lambda(x)} \over {C_n^\lambda(1)}}\right|_{\lambda=0} = T_n(x).$$ 

{\it Proof}.  We employ the $_2F_1$ form of $C_n^\lambda(x)/C_n^\lambda(1)$ given above
Proposition 1. 
Then
$$\sum_{k=0}^\infty{{(n)_k(-n)_k} \over {(1/2)_k}}
{1 \over {k!}}\left({{1-x} \over 2}\right)^k
=\cos\left(2n\sin^{-1}\left({{1-x} \over 2}\right)\right)
=\cos(n\cos^{-1}x)=T_n(x)=\left.{{C_n^\lambda(x)} \over {C_n^\lambda(1)}}\right|_{\lambda=0}.$$
\qed

{\it Remarks}.

Since $\Gamma(3/2)=\sqrt{\pi}/2$ and $J_{1/2}(x)=\sqrt{2/(\pi x)}\sin x$, it is seen that 
(3.13) is an extension of the $\lambda=1$ case of Lemma 9(a).  From Lemma 13 and (3.13) follows
the corollary that
$$\lim_{\mu \to \infty} J_\mu(x)={1 \over {\Gamma(\mu+1)}}\left({x \over 2}\right)^\mu,$$
for $|\mbox{arg} ~x| <\pi$.  This asymptotic form otherwise follows from the power series of
$J_\mu$ about the origin. 
The $\lambda \to 0$ limit for (3.13) is also more involved.  Since $J_{-1/2}(x)=\sqrt{2/(\pi x)}
\cos x$, the right side becomes $e^{xt}\cos(\sqrt{1-x^2}t)$.  That the left side agrees follows
from Lemma 15,
$$\sum_{n=0}^\infty \cos(n\cos^{-1}x){t^n \over {n!}}=e^{xt}\cos(\sqrt{1-x^2}t).$$
Thus, as $\lambda \to 0$, (3.13) becomes a generating function for the Chebyshev polynomial $T_n(x)$.

By taking specific values of $x$ in Lemma 12, including $x=0$, corollary identities follow.

The initial cases $M_0(s)$ and $M_1(s)$ from Lemma 11 are readily verified to agree with (3.2).
This Lemma also shows that $M_0$ may be factored out of the transforms, in agreement with
Proposition 4. 

Lemma 9 may be used to develop other generating functions for the transforms $M_n(s)$,
and then, with expansion in powers of $t$, other forms of $M_n(s)$ itself.

From the polynomials $p_n$, we may perform various transformations to other polynomials with zeros only on the critical line.  

Relation (2.3) may be viewed as a certain Mellin transform.

We show elsewhere \cite{coffeylettunpub} that the $\beta=1/2$ special case of Proposition 3
corresponds to a particular Mellin transform of Legendre functions.

Owing to the form of $p_n(s;\beta)$ in terms of $s(s-1)+b$, where $b$ is a rational number,
the polynomials $p_n(s;\beta_1) \pm p_n(s;\beta_2)$ also have zeros only on the critical
line, when this combination does not degenerate to a constant.  





\medskip
\centerline{\bf Certain Mellin transforms of Chebyshev polynomials $T_n$}
\medskip

The Chebyshev polynomials of the first kind begin with $T_0=1$ and $T_1(x)=x=xT_0$, and
satisfy the same recursion as $U_n$.
We now put, for Re $s>0$,
$$M_n^T(s) \equiv \int_0^1 x^{s-1} T_n(x) (1-x^2)^{1/2}dx.  \eqno(4.1)$$
For $n$ odd these also hold for Re $s>-1$.
$M_n^T(s)$ satisfies the mixed recursion relation (3.1), with
$$M_0(s)={\sqrt{\pi} \over 4}{{\Gamma\left({s \over 2}\right)} \over
{\Gamma\left({s \over 2}+{3 \over 2}\right)}}, ~~~~~~
M_1(s)=M_0(s+1)={\sqrt{\pi} \over 4}{{\Gamma\left({{s+1} \over 2}\right)} \over
{\Gamma\left({s \over 2}+2\right)}}.  \eqno(4.2)$$

We have the generating function
$$T_0(x)+2\sum_{k=1}^\infty T_k(x) t^k ={{1-t^2} \over {1-2tx+t^2}}.  \eqno(4.3)$$
{\bf Lemma 16}.  (Generating function.)  For Re $s>0$,
$$G^T(t,s) \equiv M_0(s)+2\sum_{k=1}^\infty M_k^T(s)t^k =(1-t^2)\int_0^1 {x^{s-1} \over 
{(1-2tx+t^2)}}(1-x^2)^{1/2} dx$$
$$={\sqrt{\pi} \over 4}(1-t^2)\left[{{\Gamma\left({s \over 2}\right)} \over {(1+t^2)\Gamma\left(
{s \over 2}+{3 \over 2}\right)}} ~_2F_1\left(1,{s \over 2};
{{s+3} \over 2};{{4t^2} \over {(1+t^2)^2}}\right) \right.$$
$$\left. + {{2t} \over {(1+t^2)^2}}{{\Gamma\left({{s+1} \over 2}\right)} \over {\Gamma\left({s \over 2}+2\right)}} ~_2F_1\left(1,{{s+1} \over 2};{{s+4} \over 2}; {{4t^2} \over {(1+t^2)^2}}\right) \right].  \eqno(4.4)$$

{\it Proof}.  Is similar to that for Lemma 4.

The generating function $G^T(t,s)$ may be expanded similarly as in Lemma 5.

{\bf Proposition 7}.  The polynomial factors of $M_n^T(s)$ have their zeros at the even or
odd integers up to $n-3$, and at $n^2-1$.

{\it Proof}.  This follows since when $n \geq 3$ is odd,
$$M_n^T(s)={\sqrt{\pi} \over {4\cdot 2^n}}(s-2)(s-4)\cdots(s-(n-3))(s-(n^2-1))
{{\Gamma\left({{s+1} \over 2}\right)} \over {\Gamma\left({{s+n+3} \over 2}\right)}}, \eqno(4.5a)$$
and when $n \geq 2$ is even,
$$M_n^T(s)={\sqrt{\pi} \over {4\cdot 2^n}}(s-1)(s-3)\cdots(s-(n-3))(s-(n^2-1))
{{\Gamma\left({s \over 2}\right)} \over {\Gamma\left({{s+n+3} \over 2}\right)}}. \eqno(4.5b)$$
For $n=3$ and $n=2$ only the $s-(n^2-1)$ linear factor is present.  It is then easily
verified that these expressions satisfy the recursion (3.1) with the starting functions
$M_0(s)$ and $M_1(s)$.
\qed

\medskip
\centerline{\bf Discussion}
\medskip

Our results invite several other research questions, such as: is there a combinatorial
interpretation of $p_n(s)$, and more generally, of $p_n(s;\beta)$?  Relatedly, is there a
reciprocity relation for $p_n(s)$ and $p_n(s;\beta)$?

There are several open topics surrounding the recursion (1.2).  These include:
is it possible to reduce this three-term recursion to two terms, can a pure recursion
be obtained, and, can some form of it be used to demonstrate the occurrence of the
zeros only on the critical line?

The following approach may be used to demonstrate that the $_3F_2(1)$ function in $M_n(s)$
is decomposable as a sum over $_2F_1(-1)$ functions, and to find the functional equation of
$p_n(s)$.  The Chebyshev polynomials satisfy \cite{mason} (p. 7)
$U_n(x)=U_{2n+1}(u)/(2u)$, with $u=\sqrt{1+x}/\sqrt{2}$.  Then we also have
$$M_n(s)=\sqrt{2} \int_{1/\sqrt{2}}^1 {{(2u^2-1)^{s-1}} \over {(1-u^2)^{1/4}u^{1/2}}}
U_{2n+1}(u)du.$$
The polynomial $U_{2n+1}(u)$ is odd in $u$, so that $M_n(s)$ is a finite sum of terms of the form, for Re $s>0$,
$$\int_{1/\sqrt{2}}^1 {{(2u^2-1)^{s-1}u^{2j+1/2}} \over {(1-u^2)^{1/4}}}du
=2^{-j-7/4}{{\Gamma\left({1 \over 4}-j-s\right)} \over {\Gamma\left({1 \over 4}-j\right)}}
\Gamma(s) ~_2F_1\left({1 \over 4},j+{3 \over 4};j+s+{3 \over 4};{1 \over 2}\right)$$
$$+2^{s-2}\Gamma\left({3 \over 4}\right){{\Gamma\left(-{1 \over 4}+j+s\right)} \over {\Gamma\left({1 \over 2}+j+s\right)}} ~_2F_1\left(1-s,{1 \over 2}-j-s;{5 \over 4}-j-s;{1 \over 2}\right).$$
(Here the argument $1/2$ may be transformed to either $2$ or $-1$.)  Thus, the $_3F_2(1)$
functions of Lemmas 6 and 8 are decomposable in terms of certain $_2F_1$ functions.




The Gegenbauer polynomials have the integral representation
$$C_n^\lambda(x)={1 \over \sqrt{\pi}}{{(2\lambda)_n} \over {n!}}{{\Gamma\left(\lambda+{1 \over 2}
\right)} \over {\Gamma(\lambda)}} \int_0^\pi (x+\sqrt{x^2-1}\cos \theta)^n \sin^{2\lambda-1}
\theta ~d\theta.  \eqno(5.1)$$
Then binomial expansion of part of the integrand of $M_n^\lambda(s)$ is another way to
obtain this Mellin transform explicitly.  The representation (5.1) is also convenient for
showing further special cases that reduce in terms of Chebyshev polynomials $U_n$ or
Legendre or associated Legendre polynomials $P_n^m$.  We mention as examples
$$C_n^2(x)={1 \over {2(x^2-1)}}[(n+1)xU_{n+1}(x)-(n+2)U_n(x)], \eqno(5.2)$$
and
$$C_n^{3/2}(x)={{(n+1)} \over {(x^2-1)}}[xP_{n+1}(x)-P_n(x)]=-{{P_{n+1}^1(x)} \over 
\sqrt{1-x^2}}.$$

The Gegenbauer polynomials are a special case of the two-parameter Jacobi polynomials
$P_n^{\alpha,\beta}(x)$ (e.g., \cite{andrews}) as follows:
$$C_n^\lambda(x)={{(2\lambda)_n} \over {\left(\lambda+{1 \over 2}\right)_n}} P_n^{\lambda-1/2,
\lambda-1/2}(x).$$
The Jacobi polynomials are orthogonal on $[-1,1]$ with respect to the weight function
$(1-x)^\alpha (1+x)^\beta$.  Therefore, it is also of interest to consider Mellin transforms
such as
$$M_n^{\alpha,\beta}(s)=\int_0^1 x^{s-1}P_n^{\alpha,\beta}(x)(1-x)^{\alpha/2-1/2}(1+x)^{\beta/2
-1/2}dx.$$

Finally, we return to the subject of a generalized Mellin transform, that is particularly
appropriate for polynomials and the Chebyshev and Gegenbauer functions considered herein, for 
which the
usual transform on all of $[0,\infty)$ does not exist.  For a given function $f$, we may
decompose it as separately having support, for instance, on $[0,1)$ and $[1,\infty)$,
leading to the contributions $({\cal M}_0f)(s)$ and $({\cal M}_1f)(s)$, respectively,
mentioned in the Introduction.  Then, by analytic continuation of each of these terms,
we obtain the Mellin transform of $f$ as a meromorphic function $({\cal M}f)(s)$, holding
through out the complex $s$-plane.  For further discussion of this generalized Mellin 
transform for a function locally integrable on $(0,\infty)$, \cite{bleistein} (esp. section 4.3) may be consulted.  

\bigskip
\centerline{\bf Acknowledgement}

We would like to thank Dr. J. L. Hindmarsh for his perceptive appraisal.

\pagebreak
\centerline{\bf Appendix}
\medskip

Here are collected various transformations of terminating $_3F_2(1)$ series.
Again $(a)_n$ denotes the Pochhammer symbol.

$$_3F_2(-n,a,b;c,d;1)={{(c-a)_n(d-a)_n} \over {(c)_n(d)_n}} ~_3F_2(-n,a,a+b-c-d-n+1;
a-c-n+1,a-d-n+1;1)$$
$$={{(a)_n(c+d-a-b)_n} \over {(c)_n(d)_n}} ~_3F_2(-n,c-a,d-a;1-a-n,c+d-a-b;1)$$
$$={{(c+d-a-b)_n} \over {(c)_n}} ~_3F_2(-n,d-a,d-b;d,c+d-a-b;1)$$
$$=(-1)^n{{(a)_n(b)_n} \over {(c)_n(d)_n}} ~_3F_2(-n,1-c-n,1-d-n;1-a-n,1-b-n;1)$$
$$=(-1)^n {{(d-a)_n(d-b)_n} \over {(c)_n(d)_n}} ~_3F_2(-n,1-d-n,a+b-c-d-n+1;a-d-n+1,b-d-n+1;1)$$
$$={{(c-a)_n} \over {(c)_n}} ~_3F_2(-n,a,d-b;d,a-c-n+1;1)$$
$$={{(c-a)_n(b)_n} \over {(c)_n(d)_n}} ~_3F_2(-n,d-b;1-c-n;1-b-n,a-c-n+1;1).$$

\pagebreak

\end{document}